# From Prompt Engineering to Epistemic Prompting: Prompt Trajectories as AI-Mediated Problem Framing in Science Education

Matteo Tuveri[1,2,3,*]

[1]Physics Department, University of Cagliari, Cittadella Universitaria di Monserrato, 09042 Monserrato (CA), Italy
[2]INFN, Sezione di Cagliari, Cittadella Universitaria di Monserrato, 09042 Monserrato (CA), Italy
[3]CIRD (Centre for Educational Research and Teaching Innovation) - Research in Education and Culture, University of Cagliari and Sassari, Sardinia, Italy

*Corresponding author: matteo.tuveri@unica.it

**Abstract**

Prompt engineering is commonly presented as a technical competence for obtaining more accurate, relevant, or well-formatted outputs from large language models (LLMs). However, in STEM education, prompting should also be understood as a continuous epistemic practice. Students interpret contextual and disciplinary cues and adopt expectations about what kind of knowledge, representation, and action are appropriate. Drawing on epistemological framing, and AI-mediated concept-to-decision reasoning, the paper presents a new framework called *epistemic prompting* and proposes a multi-turn Framing-Prompting Loop. The educationally relevant outcome is a prompt framing trajectory: the sequence of prompts, model responses, learner uptake, disciplinary checks, and reframing moves through which a knowledge task develops. In this framework, an initial prompt establishes a provisional macro-frame by selecting the problem, representations, assumptions, criteria, and distribution of work between learner and model. Each subsequent learner turn can then maintain, specify, challenge, repair, or transform that organization. The implications for AI-mediated STEM instruction, and, specifically, on learner-LLM interaction are also discussed.

## 1. Introduction

Large language models (LLMs) can rapidly produce explanations, summaries, examples, code, and worked solutions. Their educational value, however, cannot be inferred from fluency or task completion alone. In STEM (Science, Technology, Engineering, Mathematics) education, plausible response may conceal an inappropriate model, an unstated assumption, a mismatch between representations, or a result that lacks physical or empirical meaning. Studies in physics education have shown both the accessibility of LLM-supported activities and the fragility of model performance on disciplinary tasks (Bitzenbauer, 2023; Kortemeyer, 2023; Küchemann et al., 2023; Polverini & Gregorcic, 2024). More broadly, the educational literature identifies opportunities for feedback and dialogue alongside risks of over-reliance, reduced cognitive effort, and acceptance of unsupported claims (Fan et al., 2025; Kasneci et al., 2023; Lee et al., 2025; Lo, 2023).

Prompt engineering has become the principal practical response to this variability. Educational reviews describe prompting as a teachable competence involving the specification of goals, context, constraints, examples, roles, and output formats, together with iterative refinement (Cain, 2024; Federiakin et al., 2024; Knoth et al., 2024; Lee & Palmer, 2025; Qian, 2025). This literature establishes that prompt design matters. Its dominant evaluative logic nevertheless remains downstream: prompts are primarily judged by whether they improve the usefulness, accuracy, or conformity of model outputs. Such criteria are necessary for tool use, but they are insufficient for learning.

Conversational AI also requires a shift from isolated prompts to sequences. The first prompt is consequential because it introduces a provisional context and role structure, but framing does not end when the first response is generated. A follow-up may preserve an established model, make a hidden assumption explicit, request another representation, contest an AI-generated claim, repair a misunderstanding, redefine the task, or introduce new criteria. Recent work that examines complete learner-LLM histories has begun to identify trajectories ranging from direct delegation to iterative refinement, and has linked the quality of follow-up questions to differences in engagement and learning (Ding et al., 2026; Shao et al., 2026). These studies support a process view, but the relation between sequential prompting and disciplinary problem framing remains under-theorized.

In this paper, we deal with that relation by proposing a new framework called *epistemic prompting*. It corresponds to the deliberate construction and revision of prompts to articulate, regulate, and evaluate how a knowledge task is framed. Its distinctive contribution is not the general claim that learners plan, monitor, and revise. It is the treatment of each learner turn as a potential, publicly inspectable modification of the epistemic organization of the problem within an accumulating conversational context. The resulting unit of analysis is what we called the prompt trajectory, not the isolated prompt or final answer.

The article addresses three questions. First, what distinguishes epistemic prompting from conventional prompt engineering? Second, how can successive prompts be interpreted as framing moves rather than as undifferentiated iteration? Third, what instructional and analytical consequences follow when prompt trajectories become the focus of STEM education?

The article is designed as a conceptual model paper (Jaakkola, 2020). After synthesizing the relevant literature on prompt engineering, AI literacy, epistemological framing, metacognition, and multiple representations, it proposes a multi-turn Framing-Prompting Loop and operational categories for analysing prompt functions. The paper is intended to generate design and research implications for AI-mediated science education. The implications for instruction, assessment, and AI literacy are also discussed.

## 2. Conceptual foundations

### *2.1 Prompt engineering and AI literacy*

Prompt engineering research has identified strategies such as contextual specification, role assignment, exemplification, decomposition, constraints, and iterative refinement. In education, these strategies have been organized into competence models, curricula, and taxonomies (Cain, 2024; Chen et al., 2025; Federiakin et al., 2024; Qian, 2025). AI literacy frameworks place these operational skills alongside critical evaluation, ethics, and understanding of system limitations (Knoth et al., 2024). Prompting is therefore more than typing a question: effective use requires knowledge of the task, the system, and the criteria by which a response should be judged.

Yet, prompt quality is commonly operationalized through model performance, adherence to requested form, or user satisfaction. Even educational frameworks can implicitly treat the learner's prompt as an input variable whose function is to optimize output. This orientation obscures what prompt formulation reveals about the learner's interpretation of the task. A technically elaborate prompt may obtain an excellent answer while the learner remains unable to reconstruct its assumptions or evaluate its validity. Conversely, a prompt that produces an imperfect answer may be educationally productive when the response exposes a misconception, representational conflict, or missing criterion that becomes the object of subsequent inquiry.

Iteration is not intrinsically educational. Rephrasing a request until a desired-looking answer appears may reflect trial-and-error optimization rather than monitoring. Novice users often rely on anthropomorphic expectations and local wording changes without an adequate account of how prompts constrain generation (Zamfirescu-Pereira et al., 2023). The relevant issue is what the learner notices, preserves, contests, and validates across revisions.

Research on educational chatbots and intelligent tutoring systems has long emphasized dialogue and feedback distributed across multiple exchanges (Kuhail et al., 2023). LLM-based systems extend these possibilities because explanations and questions can be adapted locally to a learner's current formulation. Role-engineered tutors can be constrained to ask questions, interpret diagrams, or delay direct solutions, while unconfigured general-purpose models tend to provide complete answers prematurely (Kestin et al., 2025; Tufino & Gregorcic, 2025).

Recent empirical studies move beyond single-turn evaluation. Ding et al. (2026) found that learning with LLMs depended not only on the initial request but also on whether students asked substantive follow-up questions and made warranted judgments about responses. Shao et al. (2026), analysing complete programming interactions, identified trajectories that ranged from full delegation to iterative refinement and argued that prompt histories can provide a window into self-regulation and learning orientation. These findings leave a conceptual gap we try to fill in this paper: they identify interaction patterns, but do not fully specify how each learner turn reorganizes a disciplinary task or how apparently similar revisions can have opposite epistemic consequences.

### *2.2 Epistemological framing and epistemic responsibility*

Epistemological framing represents the situated interpretation of what kind of activity is taking place and what knowledge and action count as appropriate within it (Bing & Redish, 2009, 2012; Elby & Hammer, 2010; Schön, 1983). It is characterized by three interdependent actions (Pee et al., 2015): seeing, thinking, and acting. Seeing concerns how a situation is delimited, represented, and made salient. Thinking concerns the concepts, models, assumptions, and criteria through which it is interpreted. Acting concerns the operations performed in response to that interpretation. These actions are recursive. An action can reveal a representational or conceptual problem, prompting the learner to see the situation differently and reconsider the current frame.

In STEM, reasoning is a multimodal process where framing actions are tacitly or explicitly used. Learners coordinate language, symbols, diagrams, graphs, code, data displays, models, and material operations, each of which makes some meanings available while constraining others (Ainsworth, 2006; Airey & Linder, 2009; Fredlund et al., 2012; Lemke, 1998). These resources participate in disciplinary meaning-making. Equations, for example, can function as calculation templates or as representations of physical relationships (Sherin, 2001). For instance, in physics, students may approach a task as conceptual sensemaking, formula selection, algebraic execution, or consistency checking. Productive work often depends on timely shifts among these orientations rather than persistence within a single mode (Nguyen et al., 2016).

When applied to problem solving, epistemological framing is also known as problem framing. The latter describes how learners frame a physics problem by selecting and coordinating the conceptual, representational, and operational resources that make the task meaningful and solvable. In this regard, problem framing represents a useful framework to understand why possessing relevant knowledge does not guarantee productive problem solutions. Learners interpret contextual and disciplinary cues and adopt expectations about what kind of knowledge, representation, and action are appropriate (Bing & Redish, 2009; Elby & Hammer, 2010). More recently, Semiotic Problem Framing (SPF) extended the definition of problem framing by adding a semiotic layer to standard problem framing (Tuveri et al., 2026). In SPF perspective, problem solving does not begin only when a procedure is applied, but already when the learner decides what is relevant to see, which assumptions and models to use, which representations to construct, and which criteria count as valid for the solution.

In the present article, framing is used as a discipline-sensitive example of how prompts can organize both representational work and epistemic activity. Disciplines may require different frames and validation standards, but the general issue remains the same: prompting can shape how learners define the problem, coordinate representations, and justify knowledge claims.

### *2.3 AI metacognition and epistemic responsability*

In the context of Generative AI in education, a further metacognitive demand emerges. Learners must regulate not only their own activity but also the distribution of work between themselves and the model (Chang et al., 2026). Indeed, planning, monitoring, and evaluation are central to established accounts of metacognition and self-regulated learning (Flavell, 1979; Zimmerman, 2002). In LLM interaction, a diagnosis such as 'the explanation ignored uncertainty' can be externalized as a new instruction that changes the dialogue. This does not make all prompting metacognitive. Thoughtful regulation must be distinguished from repeated answer seeking, and productive offloading must be distinguished from delegation of the assumptions, standards, and validation procedures that learners should remain able to justify (Fan et al., 2025; Tankelevitch et al., 2024). Prompting can therefore externalize part of the learner's epistemic organization of a task, even though it can never provide a transparent record of thought (Chang et al., 2026).

The concept of epistemic responsibility is therefore central. LLM responses influence attention by foregrounding concepts, supplying terminology, normalizing solution pathways, and proposing next moves. Interaction is co-framed in this limited sense. However, the relation remains asymmetric. A model generates context-sensitive continuations; it does not assume responsibility for whether evidence is adequate, whether a model applies to the material world, or whether a conclusion is warranted. Epistemic prompting consequently concerns regulated delegation: the model may contribute representations, calculations, alternatives, and critiques, while normative responsibility for framing and validation remains with learners and educators.

Transitions across representations are especially relevant to prompting. Moving from a graph to a verbal explanation, from an experimental arrangement to a diagram, or from a physical model to code can expose assumptions that remain hidden within a single mode. Kapodistrias and Airey (2025) describe purposeful transformation as representational change directed by a disciplinary purpose. Epistemic prompting applies the same criterion to requests for another format: a representational transformation is productive when it makes a relation, assumption, limitation, or test more available for scrutiny, not merely when it produces additional content.

A further specification is needed when prompt trajectories are used to analyse student artefacts. The formulation of a scientifically meaningful prompt requires the learner to decide what the problem is, which variables and representations are relevant, what assumptions should be made explicit, what kind of evidence or explanation is appropriate, which operations may be delegated, and how the response will be checked. Therefore, a frame in a prompt trajectory is educationally consequential not only when it names a concept or representation, but when it shows how that resource constrains a decision, an action, or a criterion for accepting a result. Decision-focused accounts of science and engineering problem solving are useful here because they describe expert problem solving as a sequence of decisions supported by domain-specific predictive models (Price et al., 2021). Recent work in graduate physics coursework shows that many of the decisions instructors expect students to make are not systematically practised in assigned tasks (Robbins et al., 2025).

To make the conceptual synthesis explicit, Table 1 maps the main research strands on which the framework builds, their typical units of analysis, and the specific gap addressed by epistemic prompting: the interpretation of learner turns as disciplinary framing moves within an accumulating AI-mediated trajectory.

*Table 1. Adjacent research strands and the gap addressed by epistemic prompting.*

| Research strand | Primary contribution | Typical unit of analysis | Unresolved issue |
|---|---|---|---|
| Prompt engineering / AI literacy | Strategies for specifying context, roles, constraints, examples, and format. | Prompt and resulting output. | The learner's framing process and disciplinary warrant often remain secondary. |
| Conversational tutoring | Multi-turn scaffolding, feedback, role engineering, and adaptive dialogue. | Dialogue episode or tutoring outcome. | Limited account of each learner turn as a framing move in an accumulating trajectory. |
| Metacognition / self-regulation | Planning, monitoring, control, evaluation, and revision. | Learner strategy or regulatory cycle. | Does not by itself specify what is represented, delegated, or reframed in a disciplinary task. |

| Research strand | Primary contribution | Typical unit of analysis | Unresolved issue |
|---|---|---|---|
| Epistemological framing | Context-sensitive activation and coordination of knowledge and practices. | Problem-solving episode and frame shift. | Prompt trajectories and AI-mediated co-framing remain under-theorized. |
| Multiple representations / social semiotics | Coordination and transformation of language, symbols, diagrams, graphs, models, and code. | Representation or transformation across modes. | Rarely connected to sequential prompting and the distribution of epistemic labour. |

## 3. Epistemic prompting and the Framing-Prompting Loop

### *3.1 Definition and constitutive claims*

We define epistemic prompting as the deliberate construction and revision of prompts to articulate, regulate, and evaluate how a knowledge task is framed. Its primary criterion is epistemic adequacy: whether the interaction makes the relevant objects, relations, representations, assumptions, principles, uncertainties, criteria, and distribution of epistemic labour available for scrutiny. This criterion differs from output quality without being opposed to it. A correct and useful answer remains desirable, but educational value also depends on whether learners can justify how the task was constituted and why the result should be accepted.

Six claims specify the framework:

- Every prompt selects an epistemic field. Even a minimal request foregrounds some objects, relations, and criteria while leaving others implicit.
- Prompts can enact seeing, thinking, and acting framing actions: they represent what the learner treats as the situation, express an interpretation or criterion, and assign an operation to the learner or model.
- Prompts distribute epistemic labour. Asking an LLM to execute an already justified calculation differs from asking it to define the problem, choose the model, perform the calculation, and validate the conclusion.
- Prompts can connect concepts to decisions. A prompt may ask not only what a concept means, but how it constrains a representation, calculation, approximation, observational choice, or validation criterion.
- Prompting is continuous across turns. The initial prompt establishes a provisional macro-frame; later prompts can maintain, specify, challenge, repair, transform, or replace it.
- Prompt histories are partial traces. They show how learners publicly organize requests and uptake, but an absent element cannot be assumed to be absent from thought. Analysis should therefore be triangulated with disciplinary artefacts, interviews, or independent performance when empirical claims are made.

The framework's distinctive object is the relation among prompt function, accumulated context, disciplinary standard, and learner uptake. Surface wording is insufficient. 'Solve the equation' may represent unregulated delegation when it replaces problem formulation, or a bounded procedural action after the learner has constructed and justified the model. Likewise, 'explain this graph' may be answer seeking, or it may function as a targeted test of a discrepancy identified by the learner. Epistemic meaning is trajectory-dependent.

### *3.2 Prompt functions and operational indicators*

Table 2 presents a functional vocabulary for coding learner turns. The categories are not linguistic templates and are not mutually exclusive. A single prompt may specify a boundary, challenge an assumption, request a representational transformation, and redistribute work. Coding should therefore identify the dominant function in relation to the preceding response and record secondary functions when they are necessary to explain the move.

*Table 2. Functions of prompts and operational indicators in AI-mediated STEM activity.*

| Function | What changes or remains stable | Observable indicator | Illustrative prompt |
|---|---|---|---|
| Establish | Introduces the phenomenon, representations, assumptions, role distribution, and criteria. | The turn defines what is to be treated as the problem and what the model should do. | "These data may follow an exponential model. Identify the assumptions that must hold before fitting them." |
| Maintain | Preserves the current model, representation, or criterion while extending the work. | The turn explicitly retains warranted elements and requests continuation. | "Continue from this diagram using the same system boundary." |
| Specify | Makes an object, variable, boundary, uncertainty, or criterion explicit. | A previously implicit element becomes inspectable. | "Treat the container as adiabatic and mark where this assumption enters the model." |
| Challenge | Questions evidence, validity, coherence, applicability, or a hidden assumption. | The turn identifies a possible warranting problem rather than merely requesting another answer. | "What evidence supports this mechanism over the alternative?" |
| Repair | Diagnoses an inconsistency and requests a controlled revision. | The turn identifies what is wrong and what should be preserved. | "The units are inconsistent. Repair the derivation without changing the physical model." |
| Transform representation | Moves within or across verbal, visual, symbolic, graphical, material, or computational forms. | The requested transformation has an explicit disciplinary purpose. | "Plot the residuals and explain which pattern would contradict the model." |
| Reframe | Introduces another model, scale, perspective, criterion, or definition of the task. | The turn changes what counts as relevant or how the problem is constituted. | "Analyse the same data with a non-linear model and compare the residual structures." |
| Close / synthesize | Establishes the evidential basis, limits, uncertainties, and conditions for completion. | The turn requests validation and justified closure rather than a polished summary alone. | "State what is supported, what remains uncertain, and which further check is required." |

A prompt or follow-up becomes stronger when it makes explicit how a concept, representation, model, analogy, or calculation justifies a concrete decision within the task. For instance, a visual sketch, metaphor, equation, or verbal explanation is coded as epistemically productive only when it contributes to seeing what is relevant, thinking with an appropriate model, acting through a justified operation, or closing the task under stated conditions.

A framing change should be coded only when the turn alters what the learner treats as salient, meaningful, actionable, or warranting. A wording change that leaves the task organization unchanged is not, by itself, reframing. A challenge becomes epistemically consequential when it introduces a reason for doubt or a criterion for re-evaluation. A repair differs from an unrestricted regeneration because it identifies both the defect and the elements that remain warranted. Closure is justified only when the trajectory includes an explicit or inferable disciplinary check appropriate to the task, such as dimensional consistency, a limiting case, agreement across representations, empirical fit, or topological equivalence.

### *3.3 The multi-turn Framing-Prompting Loop*

Figure 1 represents how these functions unfold. Prompt 0 establishes a provisional macro-frame by representing the task, context, role, assumptions, and criteria. The LLM response introduces explanations, representations, options, and possible next moves. Learner uptake then involves interpretation: the learner may compare the contribution with prior knowledge, inspect its assumptions, test its representations, or accept it without examination. Prompt $t + 1$ operationalizes that uptake by maintaining, specifying, challenging, repairing, transforming, reframing, or closing the task. The next response changes the conversational state and begins another uptake-validation cycle.

Epistemic prompting as a multi-turn framing process
Framing actions
Seeing
perceiving and representing the situation; identifying salient features and relevant semiotic resources
Thinking
interpreting, connecting, evaluating and justifying meanings, principles and assumptions
Acting
performing, prompting, delegating, checking, revising and re-orienting the task
1 Initial task framing Prompt 0
Seeing
Represent the phenomenon, task, context, role, assumptions and criteria
2 LLM response 0
Offers explanations, representations, options and possible next moves
3 Learner uptake
Thinking
Interpret, compare, justify, question, select and validate
4 Prompt t + 1
Acting
Formulate the next move: maintain, clarify, challenge, repair or reframe
5 LLM response t + 1
The conversation state changes; a new uptake-and-validation cycle begins
Accumulated conversational context
Earlier prompts and responses remain part of the evolving frame. Every learner turn is a prompt and therefore a potential micro-framing act.
Frames
conceptual framing • representational framing • symbolic framing • visual framing • verbal framing • procedural framing • modeling framing • interpretive framing • evaluative framing

**Figure 1**. Epistemic prompting as a multi-turn framing process. Seeing, thinking, and acting are recursive framing actions that can activate different functional frames across the evolving dialogue. Adapted conceptually from Pee et al. (2015) and Tuveri et al. (2026).

Seeing, thinking, and acting are analytical distinctions rather than a fixed sequence. A prompt can compress all three, and a model response can reopen earlier phases. A generated graph may reveal an unnoticed pattern; an equation may expose a mismatch with a diagram; a counterexample may undermine the selected model. The value of the next prompt depends on how the learner interprets this altered situation. Iteration is therefore the consequence of treating prompts as epistemic acts whose effects must be monitored, not a quality criterion in itself.

The accumulated conversational context is part of the frame. Terms introduced in earlier turns acquire local meanings; assumptions may become stabilized without being repeated; and prior responses constrain how later prompts are interpreted. This accumulation supports continuity but also creates risk. An early error can be embedded in subsequent turns and gain apparent legitimacy through repetition. A model may also align with a user's suggestion because it is conversationally plausible rather than because a disciplinary check has been performed. The loop must consequently include opportunities to return to earlier assumptions, representations, and criteria.

The proposed analysis can be represented as an ordered set of turns, each coded for (a) prompt function, (b) framing action, (c) represented or delegated object, (d) concept-to-decision warrant when present, (e) disciplinary validation criterion, and (f) epistemic consequence. Such coding does not infer private cognition from text. It documents how responsibility and justification are enacted in the interaction.

## 4. Implications for STEM education and research

### *4.1 Instruction*

Prompt instruction should not be reduced to universal templates for producing “good” outputs. Templates can remind novices to state context, constraints, criteria, and desired format, but they can also become decontextualised rituals. Instructional tasks should instead require decisions that can be discussed and revised. One design begins with an underspecified problem and asks students to formulate an initial prompt, predict what the model is likely to assume, inspect the response, and write two follow-ups with distinct epistemic functions - for example, one that challenges a modelling choice and another that changes representation.

Context-rich and cooperative problem-solving traditions offer useful precedents because they delay calculation and require qualitative representation, planning, and evaluation (Heller & Hollabaugh, 1992; Heller et al., 1992). In AI-mediated versions, the model can be assigned bounded roles such as critic, verifier, counterexample generator, or Socratic interlocutor. Students should also be expected to revise that role when the task changes. A critic may be useful during model selection, whereas a constrained calculator may be appropriate after a justified equation has been established.

Representational transformations should be purposeful. Students might ask an LLM to derive a symbolic relation from a diagram, translate a mathematical result into a mechanistic explanation, generate code that implements an already justified model, or compare a graph with a verbal prediction. The value of these moves lies in whether they expose a mismatch or make a relation available for checking. Instructors can use SPF or another discipline-appropriate framework to discuss these transitions; the general principle does not depend on a single taxonomy.

### *4.2 Prompt structure as epistemic scaffolding*

A well-structured prompt can function as an epistemic scaffold when it specifies not only what the model should answer, but also how the task should be approached, which role the model should play, which actions should be avoided, which modelling assumptions must be made explicit, which concept-to-decision warrant is needed, and which criteria should govern closure. This point connects epistemic prompting with, but also differentiates it from, prompt engineering. Reviews of prompt engineering in education identify context, role assignment, constraints, examples, output format, and iterative refinement as recurrent components of effective prompting (Cain, 2024; Federiakin et al., 2024; Knoth et al., 2024; Lee & Palmer, 2025; Qian, 2025). From the perspective proposed here, these components are educationally important when they make the learner's framing decisions explicit and available for inspection.

This is the key difference between a merely detailed prompt and an epistemically structured prompt. A detailed prompt may specify more information, but still delegate the definition of the problem, the choice of model, and the criteria of validity to the LLM. An epistemically structured prompt specifies the context of inquiry, the desired distribution of labour, the permitted and non-permitted actions, the disciplinary method to be followed, and the validation criteria to be used. This claim should not be read as evidence that contextualised prompts automatically produce expert reasoning. LLMs can still generate inaccurate, incomplete, or overconfident responses, and prompt quality does not remove the need for disciplinary scrutiny. Studies on generative AI use emphasize that interaction with LLMs imposes metacognitive demands on users, who must monitor, evaluate, and control the model's contributions rather than simply request them (Tankelevitch et al., 2024). Research on non-expert prompting also shows that users may revise prompts opportunistically and unsystematically, without developing stable criteria for evaluating the resulting responses (Zamfirescu-Pereira et al., 2023, Chang et al., 2026). For this reason, the educational objective is not to teach students a universally effective prompt formula, but to help them understand why particular prompt components matter epistemically.

Prompting can therefore be understood as more than the formulation of effective instructions for a generative system. Across a trajectory, it can support the articulation and regulation of how knowledge is constructed, represented, delegated, challenged, and validated. In contexts where generated answers are increasingly accessible, the decisive competence is maintaining epistemic control over inquiry: determining what the problem is, which assumptions and representations organize it, what the model may appropriately contribute, which decision a concept justifies, and what evidence is required before accepting that contribution.

A useful instructional schema can therefore ask students to construct prompts around five elements: (a) the disciplinary context of the problem; (b) the role assigned to the model; (c) the actions the model should and should not perform; (d) the representational or methodological path to be followed; and (e) the validation criteria required before accepting a result. For example, in a physics problem-solving activity, students might be asked to write prompts that explicitly state: "Do not solve the problem immediately; first help me identify the relevant representation, then ask me which physical quantity changes, then help me connect the phenomenon with the appropriate equation, and finally require dimensional and limiting-case checks." Such a prompt is not a template to be copied mechanically. It is a compact externalization of a problem frame.

This connection is central to the proposed framework. Prompt structure is not only a technical interface for improving model behaviour; it is also a way of organizing seeing, thinking, and acting in the Framing-Prompting Loop. The context and representation specified in the prompt shape what is to be seen as relevant. The methodological instructions and conceptual questions could shape how the situation is to be interpreted. The constraints on the model's role could shape what actions are delegated and what remains the learner's responsibility. The validation criteria shape when closure is warranted. A prompt of this kind is educationally significant because it makes the epistemic architecture of the solution more visible: it delays premature formalism, preserves learner agency, and turns the model's contribution into a resource for disciplined inquiry rather than a substitute for it.

This also reframes how prompt writing should be taught. Students should not only learn to make prompts more specific; they should learn to make prompts more accountable to disciplinary standards. Instructors can compare two prompts for the same task and ask which one better specifies the context, role, method, representation, assumptions, and checks. They can also ask students to revise an LLM response by adding the missing prompt component: a boundary condition, a diagrammatic request, a conceptual interpretation, a phenomenological explanation, or an evaluation criterion. In this way, prompt design becomes directly connected to epistemic agency. The learner is not rewarded for obtaining a polished answer, but for maintaining control over how the problem is framed, how the model contributes, and why the final result should be trusted.

### *4.3 Assessment*

Assessment based only on the final prompt or answer misses how the task was organized. A complete trajectory can reveal whether a learner identifies hidden assumptions, requests warrants, responds to contradictions, coordinates representations, and retains responsibility for validation. It can also reveal unproductive patterns such as repeated answer seeking, superficial rephrasing, acceptance of mutually inconsistent responses, or unwarranted closure. Prompt histories are therefore a potentially useful form of process evidence when combined with calculations, diagrams, code, data analyses, written explanations, and independent learning measures.

A trajectory-oriented rubric could examine the adequacy of initial framing, the quality of uptake, the function of follow-up prompts, representational coordination, distribution of epistemic labour, and justification of closure. Such a rubric should not reward prompt length, rhetorical polish, or the number of iterations. A terse expert prompt may be highly productive; a long prompt may reproduce a template without understanding. Short oral questions or stimulated-recall interviews can clarify why a student preserved or changed a particular frame.

This approach also changes feedback. Rather than replacing a student's prompt with an optimized version, an instructor can ask which assumption the response introduced, what evidence would justify it, which representation is missing, what remains the learner's responsibility, and what criterion would warrant closure. These questions direct attention to epistemic regulation rather than surface formulation.

## 5. Conclusion

This conceptual article examined prompting as an epistemic practice in AI-mediated learning. Its central contribution is the shift from the isolated prompt and its output to the prompt trajectory: the evolving sequence of prompts, model responses, learner uptake, disciplinary checks, reframing moves, and, where available, downstream artefacts through which a knowledge task is organized. The proposed Framing-Prompting Loop describes this process as a recursive interaction among seeing, thinking, and acting. Within this loop, prompts can establish, maintain, specify, challenge, repair, transform, reframe, or close a task. The revised framework adds concept-to-decision warranting as a criterion for evaluating when a concept or representation actually constrains a decision, operation, approximation, or validation criterion.

The first research question asked what distinguishes epistemic prompting from conventional prompt engineering. The main distinction concerns evaluative orientation. Prompt engineering generally aims to improve the accuracy, relevance, usefulness, or formal conformity of model outputs. Epistemic prompting asks whether the interaction makes the construction and validation of the knowledge task available for scrutiny. Its central criterion is epistemic adequacy: whether assumptions, representations, models, uncertainties, relations, decision warrants, and standards of evidence are explicit enough to be examined and justified. Output quality remains important, but it does not alone establish educational value. A fluent answer may result from broad delegation, whereas an imperfect response may become productive when it exposes a conflict, unsupported assumption, missing representation, or missing validation criterion that the learner subsequently addresses.

The second research question concerned how successive prompts can be interpreted as framing moves rather than as undifferentiated iteration. A follow-up prompt becomes epistemically significant when it changes what is treated as salient, meaningful, actionable, or warranting within the accumulated conversational context.

The third research question addressed the instructional and analytical consequences of treating prompt trajectories as a focus of science education. Prompting instruction should move beyond recipes for producing improved answers and engage students in constructing, monitoring, and revising the epistemic organization of disciplinary tasks. Learners can be asked to identify a response's assumptions, missing representations, delegated operations, concept-to-decision warrants, validation criteria, and whether a revision constitutes a warranted repair. Prompt histories may also provide process-oriented assessment evidence by showing how learners respond to contradictions, coordinate representations, challenge generated claims, and justify closure. Such evidence should be interpreted alongside disciplinary artefacts, including calculations, diagrams, graphs, code, data, and independent explanations. The educational objective is regulated delegation: models may contribute alternatives, representations, calculations, questions, or critiques, while learners retain responsibility for defining the problem, justifying the model, and determining when a conclusion is warranted.

Epistemic prompting extends AI literacy by connecting operational competence with responsibility for inquiry. Learners need to understand that fluent outputs are not self-validating and that model behaviour is conditioned by prompts, prior turns, system design, and training data (Kasneci et al., 2023; Polverini & Gregorcic, 2024). More importantly, they need practice deciding which questions matter, which standards of evidence apply, and when the frame itself should be revised. Empirical work in physics suggests that substantive follow-up questioning and critical evaluation cannot be assumed simply because students have access to a conversational model (Ding et al., 2023, 2026).

Several limitations qualify these conclusions. The article proposes a conceptual framework and does not provide empirical evidence that epistemic-prompting instruction improves understanding, transfer, self-regulation, or independent problem solving. The framework also exposes an equity problem. Elaborate natural-language prompting may privilege students with stronger academic language, greater disciplinary knowledge, or prior familiarity with AI systems. Requiring polished prompts as an end in itself could conflate linguistic fluency with epistemic competence. Instruction should therefore provide multiple entry points, including spoken dialogue, sketches, equations, graphs, data tables, images, and code, and should evaluate the quality of framing across modes (Airey & Linder, 2009; Kapodistrias & Airey, 2025). Agency is preserved when learners can contest the model, redefine the task, and justify closure. It is weakened when the model silently supplies the problem definition, method, and evaluation criterion. The instructional objective is not minimal AI assistance, but visibility and control over what is delegated and what remains the learner's epistemic responsibility.

Future research should test and refine the framework across educational levels, disciplines, and interaction formats. Studies should examine whether prompt trajectories help students recover from model errors, resist premature closure, coordinate representations, connect concepts to decisions, and transfer validation practices to unaided tasks. More specifically, empirical work should test whether trajectories that include challenge, repair, and justified closure moves are associated with stronger independent validation than trajectories dominated by direct answer-seeking, and whether concept-to-decision warrants predict the quality of downstream artefacts more effectively than prompt length or number of iterations. The coding categories proposed in this article also require investigation of inter-rater reliability, construct validity, and relationships with independent measures of conceptual understanding, self-regulation, and disciplinary reasoning. Future studies should extend beyond written dialogue to multimodal prompting involving diagrams, images, graphs, equations, data, simulations, code, and student-produced artefacts. Finally, attention is needed to how epistemic prompting varies with disciplinary expertise, language proficiency, prior AI experience, and technological access, since students may produce detailed prompts without being able to evaluate the warrants of model responses, and sophisticated prompting practices may otherwise reproduce existing inequalities.

### Data availability statement

No datasets were generated or analysed for the present conceptual article. The paper develops a theoretical framework based on published literature and does not report empirical student data.

## Declaration of generative AI in scientific writing

The author used AI-assisted tools (ChatGPT-5.2) only for language editing and stylistic refinement. The conceptual development, selection and interpretation of literature, argumentation, and final revisions were carried out by the author, who takes full responsibility for the content.

## Conflict of interest

There are no known conflicts of interest associated with this publication, and there has been no significant financial support for this work that could have influenced its outcome.

## Orcid ID

M Tuveri https://orcid.org/0000-0001-5686-1713